\documentclass[aps,prb,twocolumn,groupedaddress,showpacs,floatfix]{revtex4}
\usepackage{graphicx}
\graphicspath{/Tesi/Figures/Tema7/}
\newcommand{\svar}{T \ln(t/\tau_0)}
\begin{document}
\title{Magnetic relaxation in terms of microscopic energy barriers in a model of dipolar interacting nanoparticles}
\author{\`Oscar Iglesias}
\homepage{http://www.ffn.ub.es/oscar}
\email{oscar@ffn.ub.es}
\author{Am\'{\i}lcar Labarta}
\email{amilcar@ffn.ub.es}
\affiliation{Departament de F\'{\i}sica Fonamental, Universitat de Barcelona, Diagonal 647, 08028 Barcelona, Spain}

\date{\today}

\begin{abstract}
The magnetic relaxation and hysteresis of a system of single domain particles with dipolar interactions are studied by Monte Carlo simulations. We model the system by a chain of Heisenberg classical spins with randomly oriented easy-axis and log-normal distribution of anisotropy constants interacting through dipole-dipole interactions.  
Extending the so-called $T\ln(t/\tau_0)$ method to interacting systems, we show how to relate the simulated relaxation curves to the effective energy barrier distributions responsible for the long-time relaxation. 
We find that the relaxation law changes from quasi-logarithmic to power-law when increasing the interaction strength. 
This fact is shown to be due to the appearence of an increasing number of small energy barriers caused by the reduction of the anisotropy energy barriers as the local dipolar fields increase.

\end{abstract}


\maketitle

\section{Introduction}

Long-range dipolar interactions are at the heart of the explanation of many peculiar or anomalous phenomena observed in magnetic nanostructured materials. Whereas in atomic magnetic materials the exchange interaction usually dominates over dipolar interactions, the opposite happens in many nanoscale particle or clustered magnetic systems, for which the interparticle interactions are mainly of dipolar origin. 

Among the wide variety of artificially prepared systems containing nanosized magnetic clusters, some are particularly interesting for the study of the dipolar interaction in a controlled manner. Among them, we have granular metal solids consisting of fine magnetic particles embedded in a nonmagnetic matrix, in this case, for insulating matrices (for which RKKY interactions are absent), the dipolar interaction between the granules dominates over exchange via tunneling mechanisms \cite{Luisprl02,Kleemannprb01,Alliaprb99,Alliaprb01,Denardinprb02}.
In these materials, the interactions can be tuned because the metal volume fraction and average size of the granules can be varied in a controlled way. 
Frozen ferrofluids consisting of nanosized magnetic particles dispersed in a carrier liquid have also been extensively studied  \cite{Chantrellie91,Elhilojm92,Luoprl91,Morupprl94,Morupprb95,Mamiyaprl98}. 
These are considered as experimental models of random magnet systems and, in this case, the strength of the interactions can be tuned easily by controlling the concentration of particles in the ferrofluid.

In systems with reduced dimensionality, the effects of dipolar interactions are even more relevant since they allow the existence of long-range ordered phases at low temperature \cite{Merminprl66,Brunoprl01}. 
Among two dimensional systems, we find patterned media composed by regular arrays of nanoelements \cite{Martinjm03,Skomskijpcm03} of different shapes and self-ordered magnetic arrays of nanoparticles \cite{Scheinfeinprl96,Heldprb01,SunScience00,Puntesscience01,Russierprb00}, both of potential use in ultra-high density magnetic storage. 
In this kind of materials, interparticle interactions have to be prepared with a high control over the size, shape and interparticle distances in order to minimize the interparticle interactions since they could induce demagnetization of the stored information \cite{Aignprl98,Hyndmanjm02,Sampaioprb01}.
Finally, dipolar interations has proven to be essential to elucidate the ferromagnetic order and hysteresis of one-dimensional structures such as nanostripes \cite{Shenprb97,Hauschildprb98}, monoatomic metal chains \cite{Sugawaraprb97,Gambardellanature02,Lazarovitsprb03a,Lazarovitsprb03b}, nanowires \cite{Pietzschprl00,Pietzschscience01} and others \cite{Cowburnscience00,Cowburnprb02,Etzkornprl02}.
They also play a crucial role in the quantum relaxation phenomena of molecular clusters \cite{Prokofevprl98}.

While dilute systems are well understood, experimental results for dense systems are still a matter of controversy. Some of their peculiar magnetic properties have been attributed to dipolar interactions although many of the issues are still object of debate. 
Different experimental results measuring the same physical quantities give contradictory results and theoretical explanations are many times inconclusive or unclear, in what follows we briefly outline the main subjects to be clarified.
The complexity of dipolar interactions and the frustration provided by the randomness in particle positions and anisotropy axes directions present in highly concentrated ferrofluids seem enough ingredients to create a collective glassy dynamics in these kind of systems. Experiments probing the relaxation of the thermoremanent magnetization \cite{Jonssonprl95,Mamiyaprl98,Jonssonprl98} have evidenced magnetic aging and studies of the dynamic and nonlinear susceptibilities \cite{Jonssonprl98,Djurbergprl97,Kleemannprb01} also find evidence of a critical behaviour typical of a spin-glass-like freezing. 
All these studies have attributed this collective spin-glass behaviour to dipolar interactions, although surface exchange may also be at the origin of this phenomenon. However, MC simulations of a system of interacting monodomain particles \cite{GarciaOteroprl00} show that, while the dependence of ZFC/FC curves on interaction and cooling rate are reminiscent of a spin glass transition at $T_B$, the relaxational behaviour is not in accordance with the picture of cooperative freezing.
Moreover, it is still not clear what is the dependence of the blocking temperature and remanent
magnetization with concentration, $\varepsilon$, in ferrofluids: while most experiments
\cite{Gangopadhyayie93,Chantrellbook97,Chantrellie91,Elhilojm92,OGradyie93,Luoprl91,Dormannjm99,GarciaOteroprl00}, find an increase of $T_B$ and a decrease of $M_R$ with $\varepsilon$, others
\cite{Morupprl94,Morupprb95} observe the contrary variation in similar systems.
Finally, for disordered systems, the dipolar interaction ususally diminishes the coercive
field \cite{Gangopadhyayie93,Chantrellbook97}.

The purpose of this paper is to present the results of Monte Carlo simulations of a model of a system of nanoparticles simple enough to capture the main features observed in experiments. 
In particular, we will show that the spin-glass phenomenology described above is present even in
a simple model consisting of a spin chain with dipolar interactions and disordered anisotropy easy-axes as the only ingredients. For this purpose, we present the results of simulations of the time dependendence of the magnetization for different values of the strenght of the dipolar interaction and temperatures. With the aim to establish a connection between the microscopic energy landscape of the magnetic system and the observed relaxation laws, we will present an extension of the $\svar$ scaling method to systems with dipolar interactions that allows us to extract the distribution of energy barriers and of dipolar fields responsible for the relaxation from the relaxation curves. 

%

\section{Model}  

\begin{figure}[tbp]
\centering
\includegraphics[width= 1.0\columnwidth]{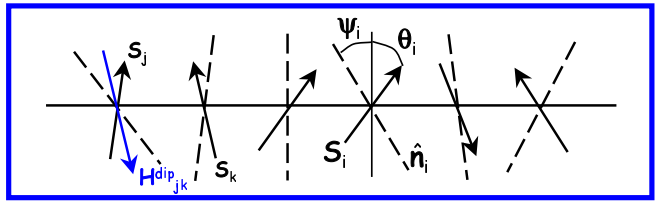}
\caption{1D chain of spins ${\bf S}_i$ with random anisotropy directions
${\bf n}_i$ (dashed lines). ${\bf H}^{dip}_{jk}$ is the dipolar field generated by the spin $S_k$ on the spin $S_j$. $\theta_i$, $\psi_i$, $\theta^{dip}$, are the angles formed by the magnetic moment, the anisotropy axis and the dipolar field with respect to the z axis.}
\label{T7_1DPart_fig}
\end{figure}

The model considered consists of a linear chain of $N = 10\,000$ classical Heisenberg spins 
${\bf S}_i$ ($i=1,\dots, N$), each one representing a monodomain particle with magnetic moment 
\mbox{\boldmath $\mu$}$_i=\mu {\bf S}_i$. As depicted in Fig. \ref{T7_1DPart_fig}, spins have 
random uniaxial anisotropy ${\bf \hat n}_i$, and anisotropy constants $K_i$, distributed according to a distribution function $f(K)$ which we will take as a lognormal
\begin{eqnarray}
	f(K)={\frac{1 }{{\sqrt{2\pi}K \sigma}}} e^{-  
\ln^2({K/K_0})/{2\sigma^2}}  \ ,
\end{eqnarray}
of width $\sigma$ and mean value $K_0$.
The spins interact via dipolar long-ranged interactions and with an external homogeneous magnetic field $\bf H$ pointing along the direction perpendicular to the chain. Spins are meant to represent the 
total magnetic moment of the particle, so that we will not take into account the internal structure of the particle.

The corresponding Hamiltonian can be written then as:
\begin{eqnarray} 
{\cal H}=-\sum_{i=1}^{N} \lbrace 
K_i ({\bf S}_i \cdot{\bf \hat n}_i)^2
+{\bf S}_i \cdot {\bf H} \rbrace \nonumber\\
+g\sum_{i=1}^{N} \sum_{j\neq i}^{N}
\left\{\frac{{\bf S}_i \cdot {\bf S}_j}{r_{ij}^3}-
3\frac{({\bf S}_i \cdot {\bf r}_{ij})({\bf S}_j \cdot {\bf r}_{ji})} 
{r_{ij}^5}\right\}
\ ,
\label{Hdip1}
\end{eqnarray}
where $g=\mu_0\mu^2/4\pi a^3$ characterizes the strength of the dipolar interaction and 
$r_{ij}$ is the distance separating spins $i$ and $j$, $a$ is the lattice spacing, 
here chosen as $1$. 
The direction of the spin vectors will be restricted to lie in the $x-z$ plane and therefore
particles are characterized by the angles $\theta_i$. This choice has been taken because only in this case exact values of the minima of the energy function and the respective energy barriers can be conputed exactly. Finally, periodic boundary conditions along the chain are considerd, so that we get rid of the possibility of spin reversal at the boundaries of the system because of the reduced coordination there. In what follows, temperature will be measured in reduced units $k_BT/K_0 V$. 

The effect of the dipolar interaction can be more easily understood by 
defining the dipolar fields acting on each spin $i$ (see Fig. \ref{T7_1DPart_fig}) 
\begin{equation}
{\bf H}^{dip}_i=-g\sum_{j\neq i}^{N} 
\left\{ \frac{{\bf S}_j}{r_{ij}^3}-3
\frac{({\bf S}_j \cdot {\bf r}_{ji}){\bf r}_{ij}}
{r_{ij}^5}\right\} \ .
\end{equation} 
Therefore, rewriting the dipolar energy as 
\begin{equation}
{\cal H}_{\rm dip}=-\sum_{i=1,}^{N}{\bf S}_i \cdot {\bf H}^{dip}_i \ ,
\end{equation}
the total energy of the system can be expressed in the simple form
\begin{equation}
{\cal H}= -\sum_{i=1}^{N} \lbrace K_i ({\bf S}_i \cdot{\bf \hat n}_i)^2-
 {\bf S}_i \cdot {\bf H}_i^{eff}) \rbrace \ .
\end{equation}

Now, the system can be thought as an ensemble of non-interacting spins feeling an effective field which is the sum of ans external and a locally changing dipolar field ${\bf H}_i^{eff}={\bf H}+ {\bf H}^{dip}_i$.
Note that the first term in Eq. (\ref{Hdip1}) is a demagnetizing term since it is
minimized when the spins are antiparallel, while the second one tends to align
the spins parallel and along the direction of the chain. For systems of aligned 
Ising spins only the first term is non-zero and, consequently, the dipolar
field tends to induce AF order along the direction of the chain (the ground state configuration for this case). 
However, for Heisenberg or planar spins, the competition between the two terms  
gives rise to frustrating interactions, which may induce other equilibrium configurations, depending on the interplay between anisotropy and dipolar energies.
\section{Computational details}
\label{Comput_details}

When considering Heisenberg spins with a continuous degree of freedom $\theta$, special care has to be taken in the way the trial steps are done \cite{Nowakann01,Hinzkecpc99}. Moreover, independently of the election of the trial step, there are different ways of implementing the Monte Carlo dynamics in this case, that differ essentially in how the energy difference $\Delta E$ appearing in the Boltzmann probability is computed.
Either $\Delta E$ is computed as the energy difference between the current 
${\bf S}^{old}$ and the attempted ${\bf S}^{new}$ values of the spin or 
it is chosen as the energy barrier which separates ${\bf S}^{old}$ 
and ${\bf S}^{new}$. Note that the second choice gives $\Delta E$'s that are higher than the first if there is an energy maximum separating the two states. Consequently, the time scale corresponding to one MC step depends crucially both on the trial step election and the chosen dynamics \cite{Nowakprl00,Nowakann01}.

Since our major interest is to study the connection between the intrinsic energy barrier distributions and the long time relaxation of the magnetization, we have devised a MC algorithm that considers trial jumps only between orientations corresponding to energy minima randomly chosen with equal probabilities. The $\Delta E$ in the transition probability are always equal to one of the actual energy barriers of the system. This is possible because in the model considered the spins are restricted to point in the x-z plane and for this case it is possible to find the energy minima and maxima as well as the energy barriers separating them numerically since the energy of a particle can be rewritten as 
\begin{equation}
E_i=-K_i \cos^{2}(\theta_i-\psi_i)-H_{i}^{eff}\cos(\theta_i^h -\theta_i)  \ ,
\label{En1}
\end{equation} 
where the $\theta_i$, $\psi_i$ and $\theta_i^h$ are the angles formed by the magnetic moment, anisotropy axis and effective field with respect to the z axis.
Although the energy barriers cannot be analytically calculated for all the values of $\psi_i$ and $\theta_i^h$, it is not difficult to build up an algorithm that finds the minima and maxima of the energy function (\ref{En1}) and their respective energies \cite{OscarThesis}. Therefore, a MC step consists of the following steps: a spin is chosen at random, the energy barriers are computed following the above mentioned method, a trial jump is attempted and accepted with probability $p_i=\exp(-\Delta E/k_B T)$ if $\Delta E>0$ or $p_i=1$ if $\Delta E<0$, the dipolar fields $H^{dip}_i$ acting on the other particles are recalculated and finally the whole process is repeated $N$ times.


\section{Relaxation curves: $\svar$ scaling with interactions}
In this section, we present the results of MC simulations of the thermal relaxation of the magnetization obtained following the protocol described in Sec. \ref{Comput_details}. The main goals are to study the variation of the relaxation law with the interaction strength $g$ and to apply $\svar$ scaling approach of the relaxation curves to show how the energy barrier distributions can be obtained from this kind of analysis even when interaction among particles is present.
\begin{figure}[tbp]
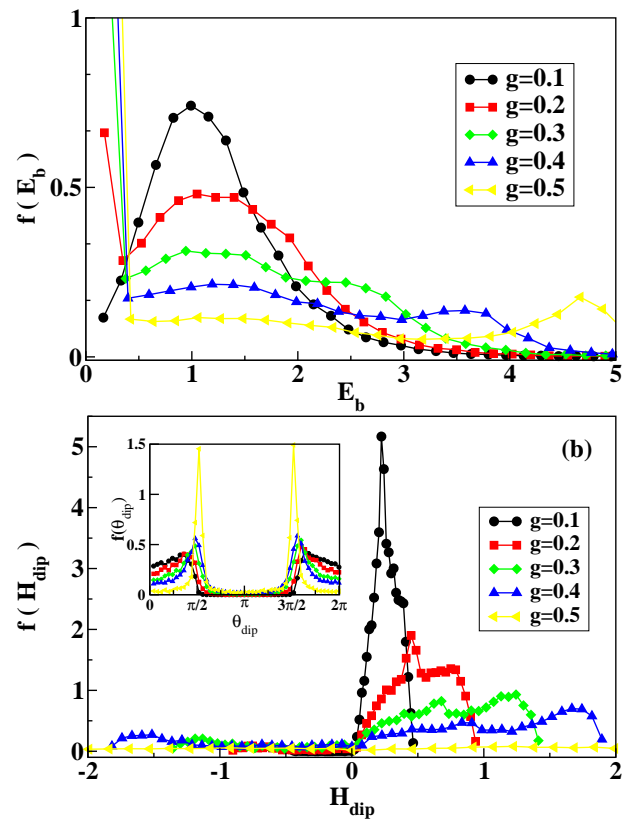

\centering
\includegraphics[width=0.45\textwidth]{Dipolar1D_fig2a.eps}
\includegraphics[width=0.45\textwidth]{Dipolar1D_fig2b.eps}
\caption{(a) Energy barrier distributions $f(E_b)$ and (b) distribution of dipolar field angles $f(\theta_{dip})$ for spin configurations achieved after an equilibration at $T=0$ in which spins have been driven itereatively towards the nearest energy minimum direction starting from an initial FM configuration.
The system has a lognormal distribution of anisotropy constants ($\sigma=0.5$) and random anisotropy axes directions.
}
\label{T7_Eb_distrib_therma_fig}
\end{figure}

\subsection{Initial configurations and effective energy barrier distributions}

The studied relaxation processes are intended to mimic experiments in which the decay of the magnetization after the application of a saturating magnetic field is recorded. Therefore, the initial spin configuration should be chosen so that all spins in the chain are pointing along the z axis. 
However, this configuration is highly metastable even at $T=0$ because, due to the randomness in anisotropy axes, the spins will not be pointing along the local energy minima directions. If the system is initially prepared in this way (by the application of a strong external field, for example), the spins will instantaneously reorient their magnetizations so that they lie along the nearest minimum. This accommodation process occurs in a time scale of the order of $\tau_0$, much shorter than the thermal over-barrier relaxation times $\tau$. Therefore, in real experiments probing magnetization at time scales of the order of $1-10$ s ({\it i.e.} SQUID magnetometry), this will not be observed. In order to get rid of this ultra-fast relaxation during the first steps of the simulations, we submit the system to a previous equilibration process at $T=0$, during which the spins are consecutively placed in the nearest energy minima. Since the dipolar field after each of this movements changes on all the spins, the energy minima positions change continuously, but, after a certain number of MC steps, the total magnetization stabilizes and the system reaches a final equilibrated state. 

The distribution energy barriers $f(E_b)$ of these initial equilibrated configurations can be obtained by sampling the individual energy barriers of all the spins using the algorithm described in Sec. \ref{Comput_details}. The normalized histograms obtained in this way are shown in Fig. \ref{T7_Eb_distrib_therma_fig} for different values of the interaction strength $g$. For weak interactions ($g= 0.1$), there are slight changes on the $f(E_b )$ with respect to the non-interacting case. As in the case of an external homogeneous field \cite{Iglesiasjap02}, the dipolar fields shift the peak of the distribution towards higher values, while its shape is unchanged. However, when increasing $g$, the smallest energy barriers of particles having the smallest $K$ start to disappear. This leads to the appearance of a peak at zero energy, to an increase in the number of low energy barriers due to the reduction by the field, and also to the appearance of a longer tail at high energies. As the dipolar interaction is increased further ($g=0.3, 0.4$), the original peak around $E_b \simeq 1$ is progressively suppressed as more barriers are destroyed, and a secondary subdistribution peaked at high energies appears as a consequence of barriers against rotation out of the effective field direction.
\begin{figure}[tbp]
\centering
\includegraphics[width= 1.0\columnwidth]{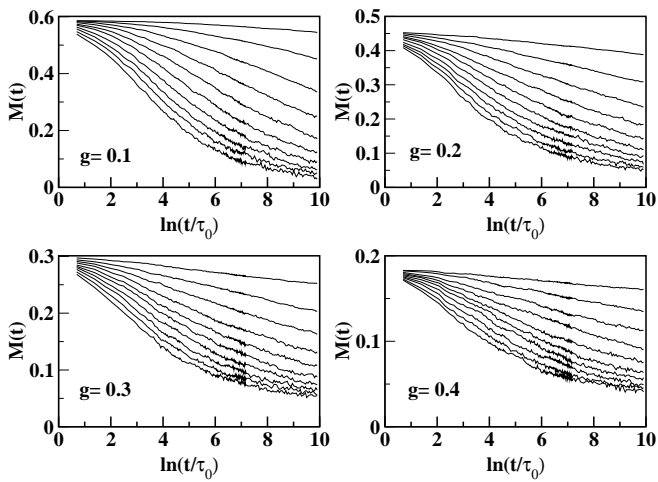}
\caption{Relaxation curves for several temperatures ranging from $T= 0.02$ (uppermost curves) to $T= 0.2$ (lowermost curves) in $0.02$ steps for a system of interacting particles with distribution of anisotropies $f(K)$ and random orientations. $g$ is the dipolar interaction strength. The initial state for all of them is the one achieved after the equilibration process described in the text.
} 
\label{T7_Relax_log_fig}
\end{figure}

\subsection{Simulations of the time dependence of the magnetization}

The relaxation curves obtained through the computational scheme described in the previous section at different temperatures are shown in Fig. \ref{T7_Relax_log_fig} for values of the interaction $g$ parameter ranging form the weak ($g=0.1$) to the strong ($g=0.5$) interaction regime.
We observe that the stronger the interaction, the smaller the magnetization of the initial configuration due to the increasing strength of the local dipolar fields that tend to depart the equilibrium directions from the direction of the anisotropy axis. Thus, we point out that, if relaxation curves for different $g$ at the same $T$ are to be compared, they have to be properly normalized by the corresponding $m(0)$ value. As it is evidenced by the logarithmic time scale used in the figure, the relaxation is slowed down by the intrinsic frustration of the interaction and the randomness of the particle orientations.

More remarkable is the fact that the stronger the interaction is, the magnetization decay is slower, which agrees well with the experimental results of Refs. \cite{Morupprl94,Batlleprb97,Garciaprb99}. 
However, at difference with other simulation works\cite{Lyberatosjpd00,Dahlbergjap94}, the quasi-logarithmic relaxation regime is only found in our simulations in the strong interaction regime, for short times, and within a narrow time window that depends on $T$. This can be understood because of the short duration of the relaxations in other works compared to ours, which were extended up to 10000 MCS, thus confirming the limitation of the logarithmic approximation to narrow time windows.
\begin{figure}[tbp]
\centering
\includegraphics[width= 1.0\columnwidth]{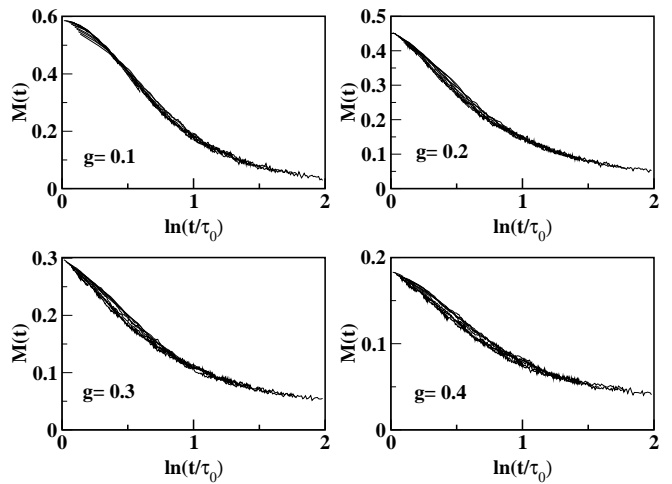}
\caption{Master relaxation curves corresponding to the relaxations shown in Fig. \ref{T7_Relax_log_fig} obtained by multiplicative scaling factor $T$. 
} 
\label{T7_Relax_log_scaling_fig}
\end{figure}

\subsection{$\svar$ scaling in presence of interaction.}
We will analyze the relaxation curves at different temperatures following the phenomenological $\svar$ scaling approach presented in previous works for non-interacting sytems \cite{Labartaprb93,Iglesiaszpb96} and systems in the presence of a magnetic field \cite{Balcellsprb97,Iglesiasjap02}. 
The method is based on the fact that the dynamics of a system of magnetic entities can be described in terms of thermal activation of the Arrhenius type over effective local energy barriers. Although one could think that this assumption is only valid in non-interacting particle systems, we would like to stress that the $\svar$ scaling approach was first successfully introduced in studies of spin-glasses, where short range frustrated interactions prevail. 
In systems with dipolar interactions, although the energy barrier landscape of the system change as the relaxation proceeds due to the long-range of the interaction, we will argue in the following sections that this fact does not preclude the applicability of scaling to low $T$ relaxations. 
In fact, the accomplishment of the $\svar$ scaling in interacting systems and 
the effective energy barrier distributions deduced from the corresponding master curves provide information about the energy barriers that are effectively probed during the relaxation process, even if they keep on changing during the process.

The results of the master curves obtained from Fig. \ref{T7_Relax_log_fig} by scaling the curves along the horizontal axis by multiplicative factors $T$, are presented in Fig. \ref{T7_Relax_log_scaling_fig} for a range of temperatures covering one order of magnitude. 
Notice that in a MC simulation $\tau_0=0.5$ and it is not an adjustable parameter of the scaling law.  
First, we observe that, in all the cases, there is a wide range of times for which overlapping of the curves is observed. 
Below the inflection point of the master curve, the overlap is better for the low $T$ curves, whereas high $T$ curves overlap only at long times above the inflection point, as in the non-interacting case \cite{Labartaprb93}. 
Moreover, it seems that scaling is accomplished over a wider range of $T$ the stronger the interaction is, whereas in the weak interaction regime, scaling is fulfilled over a narrower range of times and $T$. 
As we will explain latter, this fact is due to the different variation of the effective energy barriers contributing to the relaxation in the two regimes.  
\begin{figure}[tbp]
\centering
\includegraphics[width= 1.0\columnwidth]{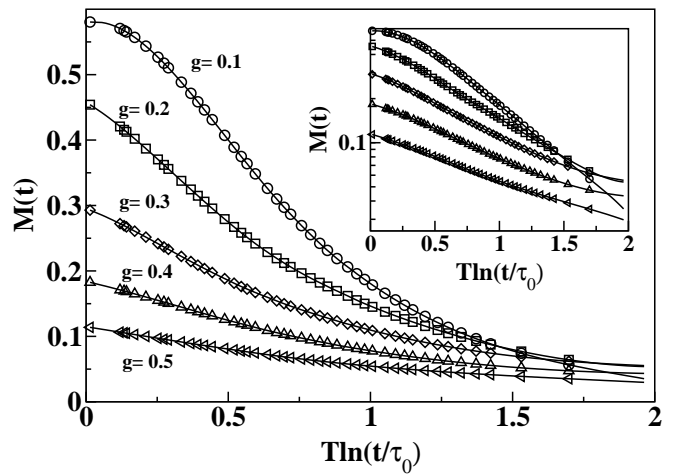}
\caption{Master relaxation curves for different values of the dipolar interaction strength $g$. Inset: the same curves in a $\ln(M)$ scale in order to evidence the power-law behaviour of the relaxation at high values of $g$.
}
\label{T7_Masters_all_fig}
\end{figure}

In order to see the influence of $g$ on the relaxation laws, we have plotted in Fig. \ref{T7_Masters_all_fig} the master relaxation curves for different values of the interaction parameter $g$ after a smoothing and filtering of the curves in Fig. \ref{T7_Relax_log_scaling_fig}. 
A qualitative change in the relaxation law can be clearly seen when increasing $g$. In the weak interaction regime ($g=0.1,0.2$), the magnetization decays to the equilibrium state with an inflection point around which the decay law is quasi-logarithmic. In the strong interaction regime ($g\ge 0.3$), however, the relaxation curves have always downward curvature with no inflection point. When plotted in a $\ln(M)$ {\it vs.} $T\ln(t/\tau_0)$ scale they are linear (see Inset of Fig. \ref{T7_Masters_all_fig}), indicating a power-law decay of the magnetization with time, since the energy scale can be converted to time through the $\svar$ variable. The curves can be fitted by $m(t)\propto t^{-\gamma}$, with a decay exponent that decreases with increasing $g$, $\gamma=1.02, 0.89, 0.74$ for $g= 0.3,0.4,0.5$ respectively. 

This power-law behaviour has also been found by Ribas et al. \cite{Ribasjap96} in a 1D model of Ising spins and by Sampaio et al. \cite{Sampaioprb01,Sampaioprb96} and Toloza et al. \cite{Tolozaprb98} in Monte Carlo simulations of the time dependence of the magnetic relaxation of 2D array of Ising spins under a reversed magnetic field. It has also been observed experimentally in arrays of micromagnetic dots tracked by focused ion beam irradiation on a Co layer with perpendicular anisotropy \cite{Hyndmanjm02,Aignprl98}, and also in discontinuous multilayers \cite{Chenprb03}.

\section{Evolution of $f_{\rm eff}(E_b)$ and of dipolar fields}
In order to gain some insight on what are the microscopic mechanisms that rule the different relaxation laws in the weak and strong interaction regimes, we will examine how the distribution of energy barriers and the distribution of dipolar fields change during the relaxation process. 
Due to the distribution of anisotropy constants and easy-axes orientations and the non-uniformity of the $T=0$ equilibrated states, it is not easy to infer the microscopic origin of the initial distributions of energy barriers shown in Fig. \ref{T7_Eb_distrib_therma_fig}a. 
It turns out that histograms of the strengh of the dipolar fields across the system for different values of $g$ turn to be useful to stablish this connection as, at low $T$, the direction and values of the local $H_{dip}$ mainly determine the first stages of the relaxation process.
Let us also notice that the distribution of dipolar fields is only sensitive to the spin orientations and their positions in the lattice and does not depend on the anisotropy or easy-axis directions of the particles, 

The computed dipolar field distibutions $f(H_{\rm dip})$ obtained by a procedure similar to that used to compute the energy barrier distributions are displayed in Fig. \ref{T7_Eb_distrib_therma_fig}b, where the dipolar fields having a component in the negative $y$ direction have been given a negative sign. 
Since most of the spins after the equilibration process are pointing along the minima closer to the positive $y$ axis, local $H_{dip}$ pointing along the negative $y$ direction will give a higher probability for the spin to jump from a metastable state to the equilibrium state. 

For weak interaction ($g= 0.1$), the initial $f(H_{\rm dip})$ is strongly peaked at a value which is very close to the dipolar field for a FM configuration $H_{\rm dip}^{\perp}=-2\zeta(3)=-2.404g$.
Dipolar fields pointing in the negative direction are scarce, indicating that the equilibrated configuration is not far from the initial FM one. 
In this case, the spins remain close to the anisotropy axis since the energy minima and the energy barriers between them do not depart appreciably from the non-interacting case. This is also corroborated by the shape of $f(E_b)$ which resembles that for $g=0$.
 
However, in the strong interaction regime, some of the local dipolar fields are strong enough to destroy the energy barriers of the particles with lower $K$, and therefore the numerous negative dipolar fields are originated by particles that have rotated into the local field direction. There are still positive fields, but now the peak due to collinear spins blurs out with increasing $g$ (it is visible at $H_{dip}\approx 0.5, 0.7$ for $g=0.2, 0.3$ respectively). At the same time, a second peak, centered at higher field values, starts to appear and finally swallows the first (see the case $g=0.5$). This last peak tends to a value equal to $H_{\rm dip}^{\parallel}= \mp\ 4.808\,g$ with increasing $g$, which corresponds to FM alignment of the spins along the chain direction. 
All these features are also supported by the distributions of dipolar field angles (see the inset in Fig. \ref{T7_Eb_distrib_therma_fig}b), which progressively peak around $\theta_{dip}=\pm \pi/2$ when increasing the interaction strength. This indicates the above mentioned tendency of spins to order along the chain direction when only one minimum is present.

\begin{figure}[tbp]
\centering
\includegraphics[width= 1.0\columnwidth]{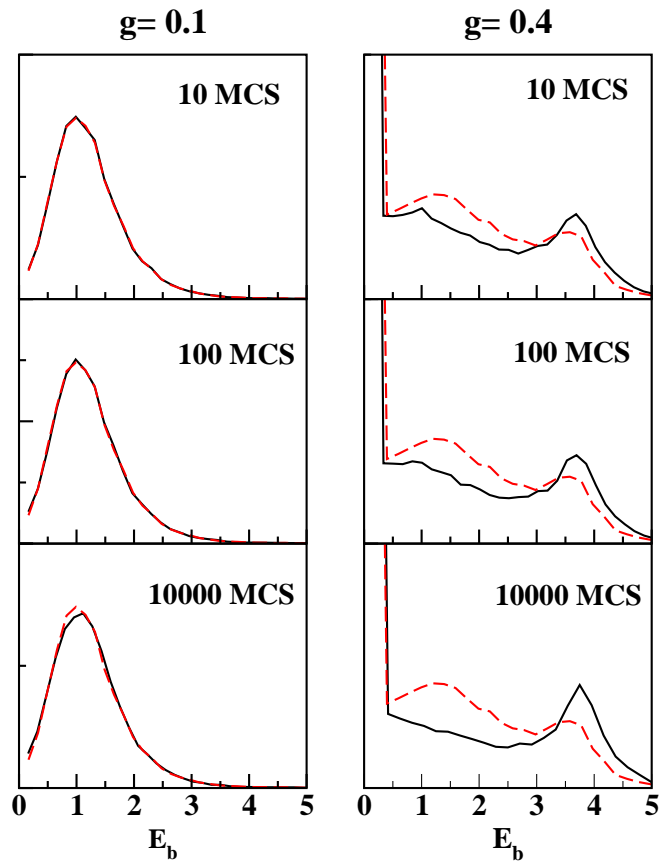}
\caption{Evolution in time of the energy barrier histograms computed at different stages of the relaxation process at $T= 0.1$. The initial distributions are shown in dashed lines.
}
\label{T7_Eb_distrib_finals_fig}
\end{figure}

In order to gain a deeper insight into the microscopic evolution of the system during the relaxation, the histograms of energy barriers and dipolar fields at intermediate stages during the relaxation have been recorded after different MC steps. 
The results for the $f(E_b)$ and $f(H_{dip})$ evolution during a relaxation at an intermediate temperature $T= 0.1$ are presented in Fig.\ref{T7_Eb_distrib_finals_fig}.
The evolutions are markedly different in the two interaction regimes.
\begin{figure}[tbp]
\centering
\includegraphics[width= 1.0\columnwidth]{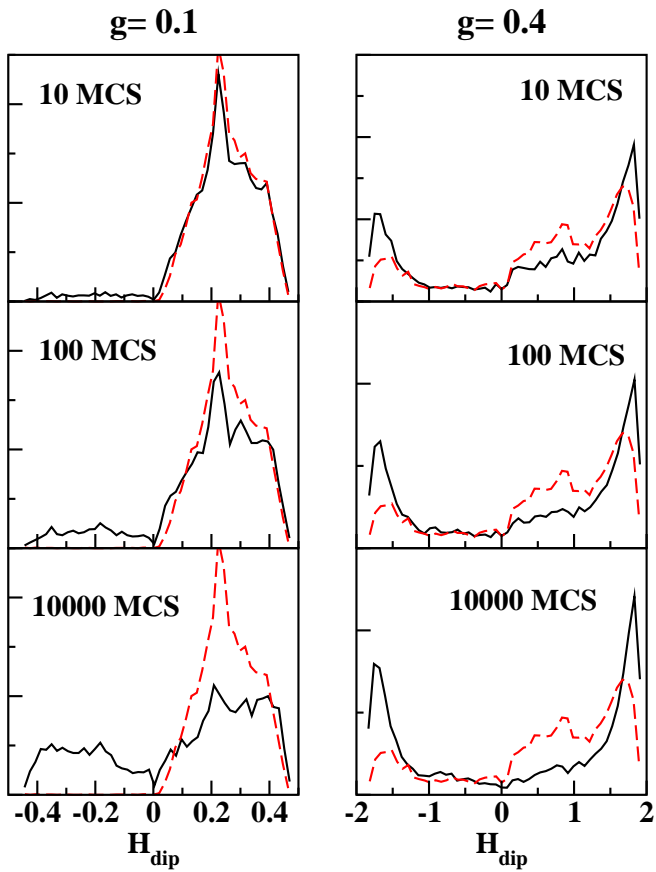}
\caption{Evolution in time of the histograms of dipolar fields computed at different stages of the relaxation process at $T= 0.1$. The initial distributions of dipolar fields are plotted in dashed lines.
}
\label{T7_Hdip_distrib_finals_fig}
\end{figure}

In the weak interaction regime, the relaxation is dominated by anisotropy barriers, so that the distributions are similar to the non-interacting case. 
As time elapses, particles with the lowest energy barriers relax towards a state with higher energy barriers. However, although during the relaxation process the energy barriers change locally, this change is compensated by the average over the anisotropy distribution and random orientations of the easy-axes. 
Thus, the global  $f(E_b)$ does not change significantly as the system relaxes, although at the final stages of the relaxation the system is in a much more disordered configuration than initially. 
In spite of this, the distribution of dipolar fields, which is more sensitive to the local changes in spin configuration, presents evident changes with time as can be seen in Fig. \ref{T7_Hdip_distrib_finals_fig}. As relaxation proceeds, the high peak of positive $H_{\rm dip}$ progressively flattens, since it corresponds to particles whose magnetization is not pointing along the equilibrium direction. 
Particles that have already relaxed, create dipolar fields in the negative direction which are reflected in a subdistribution of negative $H_{\rm dip}$ of increasing importance as time evolves. Near the equilibrium state of quasi-zero magnetization, the relative contribution of positive and negative fields tend to be equal, since, in average, there are equal number of "up" and "down" pointing spins. 

In the strong interaction regime ($g=0.4$ in Figs. \ref{T7_Eb_distrib_finals_fig} and \ref{T7_Hdip_distrib_finals_fig}), dipolar fields are stronger than anisotropy fields ($H_{\rm anis}$) for the majority of the particles, even at the earlier stages of the relaxation process. 
As time elapses, the number of small energy barriers, corresponding to the particles with smaller anisotropies, continuously diminishes as they are overcome by thermal activation. 
When relaxing to their equilibrium state, now closer to the dipolar field direction, the particles with initially small $E_b$ give rise to higher energy barriers and also higher dipolar fields on their neighbours. 
This is reflected in the increasingly higher peak in the $f(E_b)$ that practically does not relax as time elapses, causing the final distribution to be completely different from the initial one. 
What is more, as more particles relax, more particles feel an $H_{\rm dip}>H_{\rm anis}$ and, therefore, a higher $E_b$ for reversal against the local field. 
This leads to faster changes in the dipolar field distribution and also is at the origin of the power-law character of the relaxations. Equilibrium is reached when $f(H_{\rm dip})$ presents equal sharp peaked contributions from negative and positive fields, since in this case there is an equal number of particles with magnetizations with positive and negative components along the $y$ axis.

\section{Effective energy barrier distributions from $\svar$ scaling}

Our next goal is to extract the effective distributions of energy barriers from the master curves obtained from the $\svar$ scaling method and to understand what kind of microscopic information can be inferred from them in the case of interacting systems. 
In previous works \cite{Iglesiaszpb96,Balcellsprb97,Iglesiasjap02}, we have shown that in the range of validity of the $\svar$ scaling the effective distribution of energy barriers contributing to the long time relaxation process can be obtained from the master relaxation curve simply by performing its logarithmic time derivative $S(t)=dM(t)/d\ln(t)$. The obtained distribution $f_{eff}(E_b)$ represents a time independent distribution that would give rise
to a relaxation curve identical to the master curve. 
At difference from non-interacting systems (for which the $\svar$ scaling formalism was initially introduced), the $f_{eff}(E_b)$ does not necessarily match the real energy barrier distribution for the case at hand. 
\begin{figure}[tbp]
\centering
\includegraphics[width= 1.0\columnwidth]{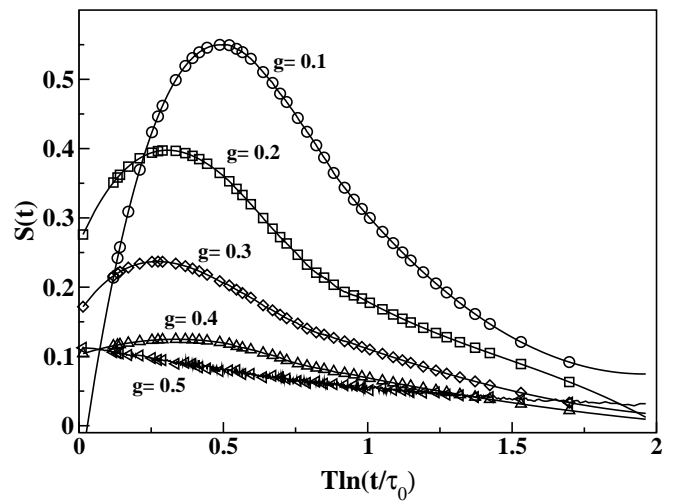}
\caption{Derivatives of the master relaxation curves of Fig. \ref{T7_Masters_all_fig} for different dipolar interaction strengths $g$.
}
\label{T7_Derivades_master_fig}
\end{figure}

Fig. \ref{T7_Derivades_master_fig} presents the $f_{eff}(E_b)$ for different values of $g$ obtained from the master curves of Fig. \ref{T7_Masters_all_fig}.
For weak interaction ($g=0.1$), the effective distribution of energy barriers has essentially the same shape as for the non-interacting case. However, as $g$ increases, the distribution becomes wider with respect to the non-interacting case and the mean effective barrier is shifted towards lower values of the scaling variable until for $g\gtrsim 0.1$ a contribution of almost zero energies dominates. In a sense, this features resemble the situation for a non-interacting particle system in an external magnetic field, for which the shift of $f_{eff}(E_b)$ with increasing $H$ is associated to the decrease of the energy barriers for rotation towards the field direction \cite{Iglesiasjap02,Iglesiascostp3}. 

When entering the strong interacting regime,
the effective distribution is clearly distorted with respect to the non-interacting case, becoming a decreasing function of the energy at high $g$. In this regime, dipolar interactions do not only modify the existing anisotropy barriers but also create high energy barriers, that result in a more uniform effective distribution spreading to higher energy values.
This change in $f_{eff}(E_b)$ is clearly related to the power-law behaviour of the relaxation law in the strong $g$ regime and, therefore, a genuine effect of the dipolar interaction. 
This striking behaviour has important consequences on the experimental interpretation of relaxation curves. A parameter oftenly used to characterize thermal contributions to magnetic relaxation is the so-called magnetic viscosity ($S(T)$ {\sl i.e.}, the slope of the magnetic relaxation curve at a given $T$ in the logarithmic dependence range). 

This change of behaviour in the effective energy barrier distributions has been observed experimentally in ensembles of Ba ferrite fine particles \cite{Batlleprb97,Batllejpd02}, in which evidence of $\svar$ scaling of the relaxation curves was demonstrated and the relevance of demagnetizing interactions in this sample was established by means of Henkel plots at different $T$. In this experiment, the authors also studied relaxation processes after different cooling fields and found that when increasing the cooling field, the effective distributions changed from a function with a maximum that extends to high enegies to a narrower distribution with a peak at much lower energy scales for high cooling fields. The effective distribution at high $H_{\rm FC}$, which was there argued to be given by the intrinsic anisotropy barriers of the particles, appears shifted towards lower energy values with respect to the anisotropy distribution as derived from TEM due to the demagnetizing dipolar fields generated by the almost aligned spin configuration induced by the $H_{\rm FC}$. From magnetic noise measurements on self-assembled lattices of Co particles, Woods et al. \cite{Woodsprl01} also extracted anisotropy energy distributions wider than nanoparticle volume distributions, an effect that can be ascribed to the strong dipolar interactions among the closely packed particle lattices. Finally, a widening of the measured barrier distributions with increasing intergranular magnetostatic interactions has been observed in a FePt nanoparticle systems \cite{Wujap03} and perpendicular media for magnetic recording \cite{Wuapl02}, which is also in agreement with the results of our simulations.   
\begin{figure}[tbp]
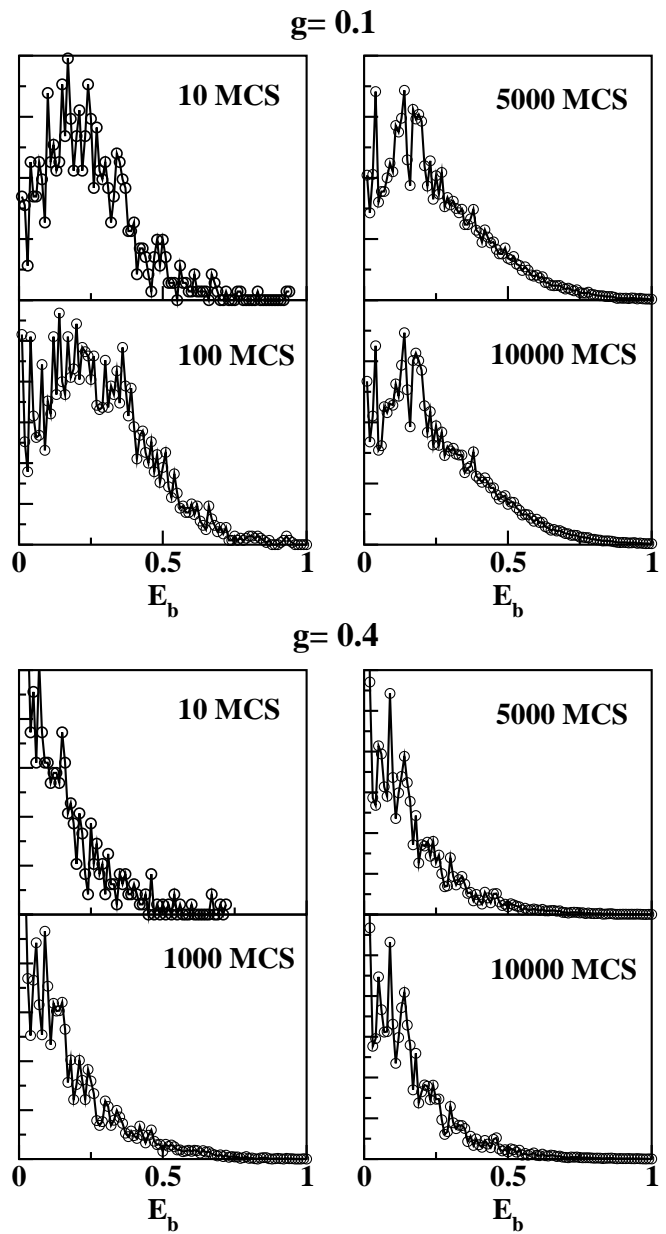

\centering
\includegraphics[width= 1.0\columnwidth]{Dipolar1D_fig9a.eps}
\includegraphics[width= 1.0\columnwidth]{Dipolar1D_fig9b.eps}
\caption{Cumulative histograms of the jumped energy barriers during the relaxation process when all the jumped energy barriers are taken into account. The temperature is $T= 0.1$. The value of the interaction parameter is $g= 0.1$ on the upper set of panels and $g= 0.4$ on the lowest set of panels.
}
\label{T7_Ebjump_g01_T010_fig}
\end{figure}
\begin{figure}[tbp]
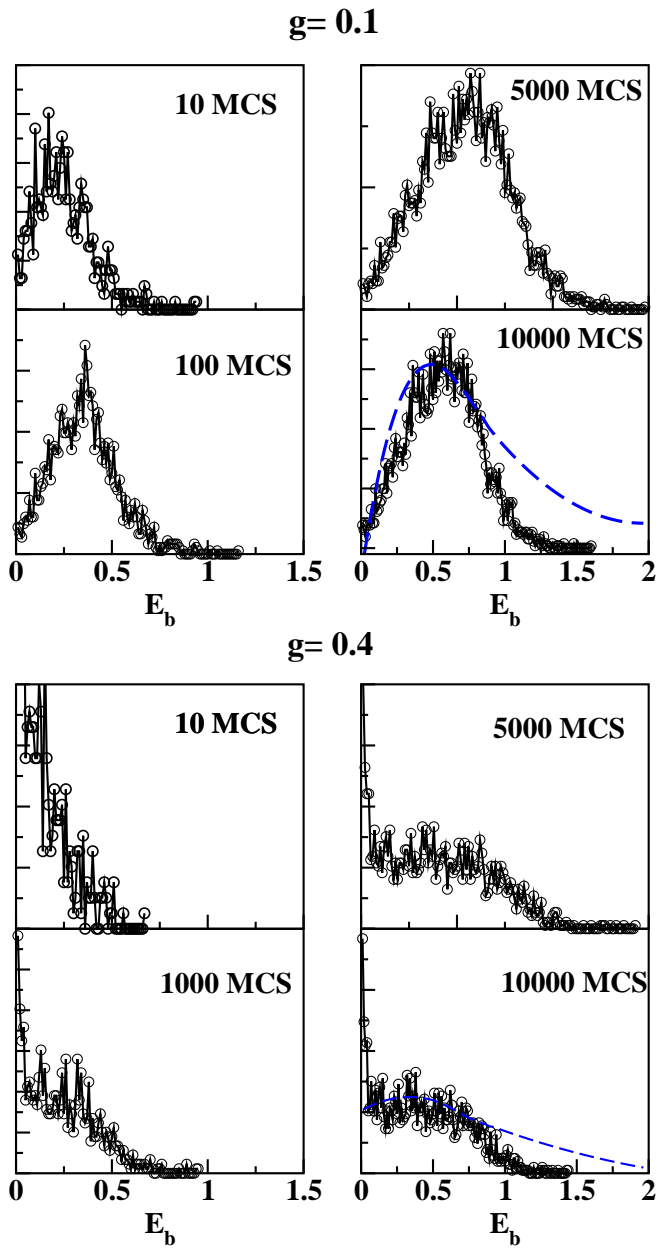

\centering
\includegraphics[width= 1.0\columnwidth]{Dipolar1D_fig10a.eps}
\includegraphics[width= 1.0\columnwidth]{Dipolar1D_fig10b.eps}
\caption{Cumulative histograms of the jumped energy barriers during the relaxation process when only the $E_b$ jumped by particles that have not jumped up to time $t$ are taken into account. Symbols correpond to $T= 0.1$. The dashed lines stand for the derivatives of the master relaxation curves shown in Fig. \ref{T7_Derivades_master_fig}. 
The value of the interaction parameter is $g= 0.1$ on the upper set of panels and $g= 0.4$ on the lowest set of panels.
}
\label{T7_Ebjumpnou_g01_T010_fig}
\end{figure}

By direct comparison of curves in Fig. \ref{T7_Masters_all_fig} with those in Fig. \ref{T7_Eb_distrib_therma_fig}, it is clear that the effective energy barrier distributions derived form the master relaxation curves do not coincide with the real energy barrier distributions.
In order to unveil the information given by $f_{eff}(E_b)$, we have computed the cumulative histograms of energy barriers that have been really jumped during the relaxation process. 
The corresponding results are presented in Fig. \ref{T7_Ebjump_g01_T010_fig} for systems in the weak and strong interaction regimes and $T= 0.1, 0.2$. 
Although in principle one could think that the derivative of the master curves collects jumped energy barriers of the order of $\svar$ as time elapses, direct comparison of 
the curves in Fig. \ref{T7_Ebjump_g01_T010_fig} with those in Fig. \ref{T7_Derivades_master_fig}, reveals that the cumulative histograms overcount the number of small energy barriers at all the studied $T$ and $g$. 
This small energy barriers that are not seen by the relaxation correspond to the those jumped by the superparamagnetic (SP) particles, which are not blocked. 

In fact, when the cumulative histograms are computed by counting only the $E_b$ jumped by particles that have not jumped up to a given time $t$ (blocked particles), the contribution of SP particles that have already relaxed to the equilibrium state is no longer taken into account. 
The histograms computed in this way are presented in Fig. \ref{T7_Ebjumpnou_g01_T010_fig}. Here, we see that when only the energy barriers jumped by the blocked particles are taken into account, the resulting histograms at advanced stages of the relaxation process tend to the effective energy barriers derived from the master relaxation curves (dashed lines in the panels at $t = 10000$ MCS). The difference between both quantities at high energy values is due to the existence of very high energy barriers, that can only be surmounted at temperatures higher than those considered here or at very long times. 

\section{Conclusions and Discussion}


We have studied the magnetic relaxation of a simple model consisting of a spin chain with dipolar interactions, showing that they are responsible for the long time dependence of the magnetization observed in many experiments. As the strength of dipolar interactions $g$ is increased, the relaxation law changes from quasi-logarithmic to a power law as $g$ increases, due to the intrinsic disorder of the system and the frustration induced by the dipolar interactions. 
This power-law decay has been observed in relaxation experiments under a reversal field of increasing magnitude in arrays of magnetic dots produced by high fluence ion irradiation \cite{Sampaioprb01,Hyndmanjm02}. MC simulations mimicking the experiments demonstrated \cite{Sampaioprb96,Sampaioprb01} that this was due to the long-range character of the interaction. 
Recent studies on granular multilayers \cite{Chenprb03} have also revealed power-law relaxations of the thermoremanent magnetization for nominal thicknesses of the magnetic layer $t_n \geq 1.2$ nm, for which superferromagnetic behaviour order between magnetic clusters was observed \cite{Sahooprb03}. In this case, the power-law behaviour was attributed to the relaxation of superspins with random anisotropy axes and distribution of anisotropies inside domains, towards more perfect collinearity. Moreover, they found a finite residual magnetization at long times that is also observed in our relaxation curves (see Figs. \ref{T7_Relax_log_fig} and \ref{T7_Relax_log_scaling_fig}) as a consequence of the competition between the randomness in anisotropy axis orientations and the frustration induced by the dipolar interactions.
Another simulation work have observed power-law decays of the magnetization in systems of ferromagnetic nanoparticles with dipolar interactions at high concentrations \cite{Ulrichprb03} that approached a finite remanent magnetization. The authors explained these result by assuming that the relaxation rate followed a power-law decay with time. Here, instead, we have been able to deduce directly the distribution of energy barriers responsible for this spin-glass like time dependence and to see that it coincides with the distribution deduced from the master relaxation curve. This energy barrier distribution is broadened and has an increasing contribution of small energy barriers as the dipolar interaction $g$ increases. These two features are in accordance with the experimentally observed broadening of the relaxation rates with respect to noninteracting particle systems found in relaxation experiments on nanosized maghemite particles \cite{Jonssonprb98} and granular multilayers \cite{Sahooprb03}.

Although our results have been obtained for a one dimensional chain of magnetic entities, we believe that similar conclusions can be drawn for systems with higher effective dimensionality as long as their magnetic behaviour is dictated by long-range interactions.

We have proved that, in the scope of our model, the $\svar$ scaling phenomenological model presented in previous works for non-interacting \cite{Labartaprb93,Iglesiaszpb96} systems and for systems relaxing in the presence of a magnetic field \cite{Balcellsprb97,Iglesiasjap02} is also valid for interacting systems within limits similar as those for non-interacting systems.  
From the master relaxation curves obtained by the application of this method, we have shown that effective energy barrier distributions can be obtained, giving valuable information about the microscopic energy barriers responsible for the relaxation. Moreover, with this method, the variation of these energy barrier distributions can be monitored as a function of the dipolar interaction strength, an information that cannot be directly measured. 

For weak interactions (diluted systems), the effective energy barrier distributions shift towards lower $E_b$ values with respect to the non-interacting case and become wider as the strength of the dipolar interaction $g$ increases, in qualitative agreement with experimental results. However, for strong interactions (dense systems), the energy barrier distributions become a decreasing function of energy with an increasing contribution of quasi-zero barriers as $g$ increases. 
We believe that both behaviours can reconcile the contradictory explanations \cite{Dormannjm99,Hansenprl03,Luisprl03} given to account for the variation of the blocking temperature $T_B$ with particle concentration in terms of energy barrier models. 
For weakly interacting systems, the energy barriers relevant to the observation time window decrease with increasing interaction and consequently the same behaviour is expected for $T_B$.
This is corresponds to the observations by M{\o}rup and collegues \cite{Morupprl94,Hansenjm98} in M\"ossbauer experiments on maghemite nanoparticles.

However, when interparticle interactions are strong enough to dominate over the disorder induced by the distribution of anisotropy axes, we have shown that the dynamic effects are ruled out by an effective energy distribution that broadens towards higher energies as $g$ increases. 
Consequently, an increase in the blocking temperature is expected as observed in the 
ac susceptibility measurements on Co clusters by Luis et al. \cite{Luisprl03}. 

\begin{acknowledgments}
  Financial support of the Spanish CICYT through the MAT2000-0858
  project and the Generalitat de Catalunya through the 2000SGR00025
  project are gratefully recognized.
\end{acknowledgments}


\end{document}